\documentclass[aps,prb,twocolumn,superscriptaddress,nofootinbib,longbibliography]{revtex4-2}
\usepackage{amsmath,amssymb,bm,graphicx,braket}
\usepackage{xcolor}
\usepackage[colorlinks=true,linkcolor=blue,citecolor=blue,urlcolor=blue]{hyperref}
\usepackage[left]{lineno}

\begin{document}
%\linenumbers  
\title{Spin-Locked Helical Currents and Charge-Neutral Spin-Channel Pumping in Altermagnetic Nanotubes}

\author{Xin Chen}
\email{xin.chen@iat.cn}
\affiliation{Thermal Science Research Center, Shandong Institute of Advanced Technology, Jinan 250100, Shandong, China}

\author{Zhen Han}
\affiliation{Thermal Science Research Center, Shandong Institute of Advanced Technology, Jinan 250100, Shandong, China}

\author{Linyang Li}
\email{linyang.li@hebut.edu.cn}
\affiliation{School of Science, Hebei University of Technology, Tianjin 300401, China}

\author{Mingwen Zhao}
\email{zmw@sdu.edu.cn}
\affiliation{School of Physics, Shandong University, Jinan 250100, China}

\date{\today}

\begin{abstract}
Altermagnetism has been widely explored in 3D and 2D crystals, but its one-dimensional realization remains largely unexplored. Here we propose an altermagnetic nanotube formed by rolling a 2D altermagnet, which converts \textcolor{black}{symmetry-enforced directional spin anisotropy in momentum space} into \textcolor{black}{spin--chirality locking in the screw-symmetric geometry}. Unlike curvature-induced magnetization in bent films, \textcolor{black}{the nanotube remains compensated and produces no net magnetization}. Two reciprocal effects emerge: (i) \textcolor{black}{spin-selective} injection drives a helical current whose handedness is fixed by the spin, yielding opposite-sign axial magnetic fields; and (ii) a time-varying axial flux generates a circumferential Faraday field that drives \textcolor{black}{equal and opposite axial charge currents in the two fixed spin channels, yielding a charge-neutral spin-channel current in the open-circuit weak-SOC limit}. \textcolor{black}{Explicit the first-principles calculations of the relaxed V$_2$Se$_2$O nanotube reveal spin-resolved helical $|\psi|^2$ modulations near both band edges, while a parent-monolayer Wannier--Boltzmann calculation quantifies the energy-dependent spin-odd transverse response and corresponding ideal thin-wall field scale.}
\end{abstract}

\maketitle
\section{Introduction}
Over the past decades, nanotubes have provided a fertile platform for exploring how curvature and confinement reshape quantum and chemical behavior\cite{AjikiAndoJPSJ1993,PhysRevLett.74.1123,doi:10.1126/science.275.5297.187}. Carbon nanotubes (CNTs), as archetypal one-dimensional conductors and semiconductors, have been extensively investigated for their helical subbands, Aharonov--Bohm flux sensitivity, and remarkable current-carrying and catalytic capabilities \cite{Bachtold1999,PhysRevLett.93.216803,doi:10.1126/science.1096524,doi:10.1126/science.275.5297.187,Yang2020ChiralityPureCNTs,doi:10.1021/acsnano.6b06900}. Recently, CNTs also serve as versatile templates and structural backbones for rolling or depositing other two-dimensional (2D) materials into tubular forms. This template-assisted route has enabled the controlled synthesis of transition-metal dichalcogenide (TMD) nanotubes and one-dimensional van der Waals heterostructures \cite{doi:10.1126/science.1096524,doi:10.1021/acsnano.0c10586,doi:10.1021/acsnano.5c11180,https://doi.org/10.1002/smll.202570208,https://doi.org/10.1002/adma.202306631}, where curvature-induced strain and symmetry reduction tune electronic, optical, and catalytic responses. 

A nanotube can be regarded as a curved two-dimensional surface. Although it appears as a one-dimensional material and is often treated as such, it still retains a degree of freedom along the circumferential direction, which gives rise to a wealth of physical phenomena. For example, many nanotubes possess structurally defined chiralities that leave clear helical fingerprints in their transport, optical, and chemical responses\cite{doi:10.1021/acsnano.6b06900}. Beyond lattice chirality, rolling magnetic two-dimensional materials into nanotubes introduces spin as an additional internal degree of freedom, thereby enabling spin‑dependent chirality under specific magnetic configurations.
\textcolor{black}{Here, spin--chirality locking means that reversing the fixed spin channel reverses the azimuthal component of the helical current while the spin-quantization axis remains fixed.}

Altermagnets are collinear magnets with vanishing net magnetization yet sizable, nonrelativistic spin splitting in momentum space \cite{PhysRevX.12.040501,PhysRevX.12.031042,PhysRevB.102.014422,smejkal2020crystal,PhysRevLett.132.236701,PhysRevX.12.011028,PhysRevLett.128.197202,Jungwirth2025,Zhu2024nature,ChenX2025,chen2025neelvectoror,10.1063/5.0147450}. \textcolor{black}{Altermagnets exhibit momentum-dependent, direction-selective splitting between two fixed spin channels. We roll such a layer into a compensated screw-symmetric nanotube whose channels are related by spin-space symmetry in the weak-SOC limit.} This mechanism is distinct from the curvature-induced magnetization reported for bent altermagnetic films \cite{PhysRevLett.134.116701}.

The central mechanism has two reciprocal forms. In the direct configuration, \textcolor{black}{spin-selective injection} produces a helical charge flow whose handedness and axial magnetic field reverse \textcolor{black}{when the selected channel is reversed}. In the inverse configuration, a time-varying axial flux drives equal and opposite axial charge currents \textcolor{black}{in the two fixed spin channels, producing a charge-neutral spin-channel current under open-circuit weak-SOC conditions}. As a related implication, spin accumulation renders the nanotube's handedness programmable and allows its spin-selected chirality to be imprinted onto nominally achiral coaxial nanotubes via one-dimensional van der Waals heterostructures, with the induced handedness reversing when the injected spin is reversed. \textcolor{black}{Using V$_2$Se$_2$O \cite{Manc2021} as a prototype, we combine relaxed-tube with a spin-resolved Wannier--Boltzmann calculation of the parent monolayer to quantify the energy-dependent spin-odd transverse response and corresponding ideal thin-wall field scale.}

\begin{figure*}[t]
\includegraphics[scale=1.0]{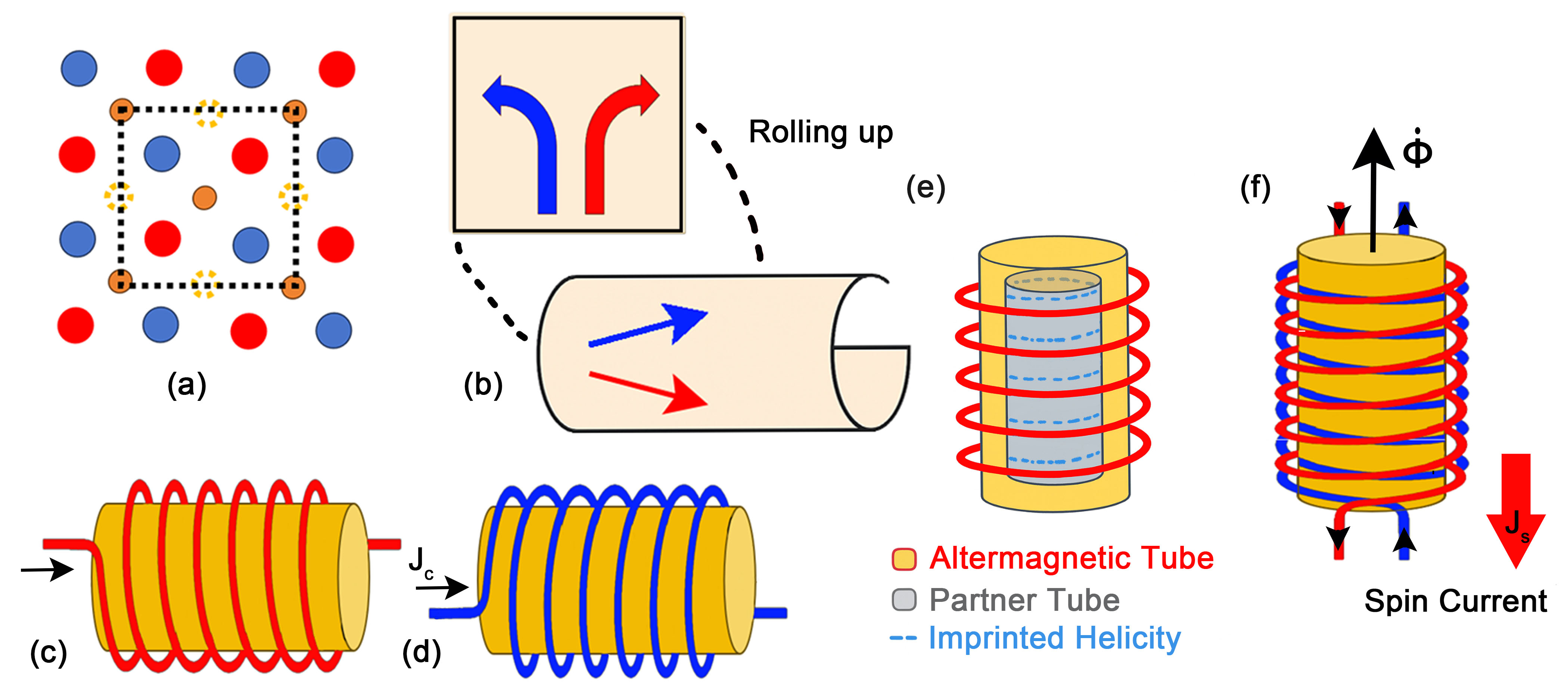}
\caption{\textcolor{black}{Geometric mapping of a spin-selective transverse response onto an altermagnetic nanotube.}
\textcolor{black}{(a),(b) Rolling maps a parent-layer transverse direction onto the tube azimuthal direction.}
(c),(d) Direct effect: \textcolor{black}{spin-selective injection} produces a helical charge current and an axial field whose sign reverses with spin.
(e) Spin-programmable chiral imprinting in coaxial altermagnetic heteronanotubes. (f) Inverse effect: a time-varying axial flux drives equal and opposite axial currents \textcolor{black}{in the two fixed spin channels, producing a charge-neutral spin-channel current in the weak-SOC limit.}}
\label{figure1}
\end{figure*}

\section{Computational details}
We perform first-principles calculations within density functional theory (DFT) as implemented in the VASP Code \cite{002230939500355X,PhysRevB.54.11169}. The exchange-correlation energy was described using the generalized-gradient approximation with an on-site Hubbard correction (PBE+U) \cite{PhysRevLett.77.3865} applied to the V-$d$ orbitals, with an effective $U_{\mathrm{eff}} = 4.3$~eV. The plane-wave basis was truncated at a kinetic-energy cutoff of 520~eV. Brillouin-zone integrations were performed using a Monkhorst-Pack mesh with a reciprocal-space resolution finer than $2\pi\times0.04$~\AA$^{-1}$. The electronic self-consistency threshold was set to $10^{-5}$~eV, and atomic positions were relaxed until the residual forces were below $0.01$~eV/\AA. \textcolor{black}{Band-structure post-processing was performed with \textsc{VASPKIT}~\cite{wang2019vaspkit}, and atomic structures were visualized with \textsc{VESTA}~\cite{Mommako5060}.}

{\color{black}
Unless stated otherwise, the electronic structures discussed below are collinear and nonrelativistic. This convention keeps the two spin channels well defined and is the limit in which the spin-space-symmetry argument in Sec.~\ref{subsec:symmetry} applies most directly. Figures~\ref{figure2}--\ref{figure4} are obtained from an explicit relaxed [110], $N=7$ V$_2$Se$_2$O nanotube calculation. The nanotube contains 14 parent cells and has a relaxed axial period of 5.772~\AA. \textcolor{black}{The parent-layer Wannier Hamiltonian was constructed with \textsc{Wannier90}~\cite{MOSTOFI20142309} using 19 functions per spin, initialized from V-$d$, Se-$p$, and O-$p$ projections on a $15\times15\times1$ full-Brillouin-zone mesh.}
}

{\color{black}
For the two-dimensional parent-layer benchmark, we evaluated the spin-resolved sheet-conductivity tensor from the Wannier Hamiltonian in the constant-relaxation-time Boltzmann approximation~\cite{Pizzi2014BoltzWann,Manc2021},
\begin{equation}
\begin{aligned}
\sigma_{ij}^{\sigma}(\mu,T)
={}&e^2\tau\sum_n\int_{\rm BZ}\frac{d^2k}{(2\pi)^2}
v_{ni}^{\sigma}(\mathbf{k})v_{nj}^{\sigma}(\mathbf{k})\\
&\times\textcolor{black}{\left[-\frac{\partial f(E;\mu,T)}{\partial E}\right]_{E=E_{n\mathbf{k}}^{\sigma}}}.
\end{aligned}
\label{eq:boltzmann-tensor}
\end{equation}
where \(\sigma=\uparrow,\downarrow\), \(i,j\in\{z,\theta\}\), \(\hat{\mathbf z}\parallel\mathbf a_1+\mathbf a_2\), and \(\hat{\boldsymbol\theta}\parallel\mathbf a_1-\mathbf a_2\). \textcolor{black}{Here \(e>0\), \(\tau\), \(\mu\), \(T\), and \(f\) denote the elementary charge, the common band-, momentum-, and spin-independent relaxation time, chemical potential, temperature, and Fermi--Dirac distribution, respectively. \(E^\sigma_{n\mathbf{k}}\) and \(|u^\sigma_{n\mathbf{k}}\rangle\) are the eigenvalue and eigenstate of the spin-resolved Wannier Hamiltonian \(H^\sigma(\mathbf{k})\). The band velocity was evaluated as~\cite{Yates2007WannierInterpolation,Pizzi2014BoltzWann}}
\begin{equation}
v_{ni}^{\sigma}(\mathbf{k})
=\frac{1}{\hbar}
\left\langle u_{n\mathbf{k}}^{\sigma}\left|
\frac{\partial H^{\sigma}(\mathbf{k})}{\partial k_i}
\right|u_{n\mathbf{k}}^{\sigma}\right\rangle .
\label{eq:wannier-velocity}
\end{equation}
\textcolor{black}{Because Eq.~\eqref{eq:boltzmann-tensor} is a two-dimensional integral, \(\sigma_{ij}^{\sigma}\) is a sheet conductance with units of siemens. No effective layer thickness is introduced. Multiplication by an in-plane electric field therefore gives a sheet current \(\mathcal K_i^\sigma\) with units of A\,m\(^{-1}\).}
We define the single-channel, spin-odd, and residual charge conversion ratios as
\begin{align}
\eta_{\sigma}(\mu)&=
\frac{\sigma_{\theta z}^{\sigma}(\mu)}
{\sigma_{zz}^{\sigma}(\mu)},\nonumber\\
\eta_{\rm spin}(\mu)&=
\frac{\sigma_{\theta z}^{\uparrow}(\mu)
-\sigma_{\theta z}^{\downarrow}(\mu)}
{\sigma_{zz}^{\uparrow}(\mu)
+\sigma_{zz}^{\downarrow}(\mu)},\label{eq:eta-spin-definition}\\
\eta_{\rm charge}(\mu)&=
\frac{\sigma_{\theta z}^{\uparrow}(\mu)
+\sigma_{\theta z}^{\downarrow}(\mu)}
{\sigma_{zz}^{\uparrow}(\mu)
+\sigma_{zz}^{\downarrow}(\mu)}.\nonumber
\end{align}
A common spin-independent \(\tau\) cancels from these dimensionless ratios. The integrations were performed at 150~K on a uniform \(300\times300\) mesh covering the complete two-dimensional Brillouin zone without symmetry reduction. Equation~\eqref{eq:boltzmann-tensor} contains the symmetric intraband product \(v_i v_j\), so the calculation evaluates the Boltzmann contribution \(\sigma_{\theta z}^{\sigma}=\sigma_{z\theta}^{\sigma}\). The dashed band-edge limits in Fig.~\ref{fig:v2-parent-transport-energy} were obtained independently from the Wannier-interpolated curvatures at the \(X\) and \(Y\) valleys.
}

{\color{black}
The band-edge masses were obtained from the curvature of the Wannier-interpolated bands~\cite{Yates2007WannierInterpolation}, evaluated using the three-point central difference
\begin{equation}
\frac{1}{|m_i^*|}
=\frac{1}{\hbar^2}
\left|
\frac{E(\mathbf k_0+\Delta k_i\hat{\mathbf i})
-2E(\mathbf k_0)
+E(\mathbf k_0-\Delta k_i\hat{\mathbf i})}
{(\Delta k_i)^2}
\right|.
\label{eq:effective-mass}
\end{equation}
We tracked the same Wannier band branch at the three points and used $\Delta k_i=7.72\times10^{-4}$~\AA$^{-1}$, corresponding to a reduced-coordinate step of $5.0\times10^{-4}$. The reported positive hole masses are local finite-difference estimates at the VBM.
}

\section{Results and discussion}

\subsection{\texorpdfstring{Symmetry mechanism of the rolled altermagnet}{Symmetry mechanism of the rolled altermagnet}}\label{subsec:symmetry}

Fig.~\ref{figure1}(a) shows a 2D Lieb-type decorated supercell representing an altermagnet \cite{Manc2021,10.1063/5.0147450,Xu2025Alterpiezoresponse,10.1063/5.0242426,v38b-5by1,PhysRevB.111.134429,10.1063/5.0252374,Khan2025}. \textcolor{black}{In the nonrelativistic limit, the two spin sublattices are related by the spin-space operation $[C_{2}\parallel C_{4z}]$. Its real-space part maps $(k_x,k_y)\to(k_y,-k_x)$, whereas its spin part exchanges the two fixed spin channels.} In this work, we adopt the Lieb-type decorated lattice as a representative 2D altermagnet.

Along the [110] direction of the Lieb-type lattice, as shown in Fig.~\ref{figure1}(b), electrons with opposite spins experience opposite transverse deflections due to \textcolor{black}{directional spin anisotropy in momentum space}. When the layer is rolled into a nanotube along the [110] direction, \textcolor{black}{the two in-plane parent directions become crossed helical directions on the cylindrical surface, converting the directional spin anisotropy into an axial--azimuthal correspondence}. \textcolor{black}{The rolled tube remains compensated, and the spin-space operation $[C_{2}\parallel M_z]$ relates its two fixed-spin sectors.}

The same symmetry logic also explains why the ideal one-dimensional dispersions can remain spin-degenerate while their wave functions carry opposite helical characters. Let $\sigma=\uparrow,\downarrow$ denote the fixed spin channel.
{\color{black}
The universal spin-only operation of a collinear magnet, $[\overline{C}_{2}\parallel\mathcal{T}]$~\cite{PhysRevX.12.031042}, gives
\begin{equation}
E_\sigma(k_z)=E_\sigma(-k_z),
\label{eq:S_spinonly}
\end{equation}
where $k_z$ is the tube-axis wave vector. The rolled-tube operation $[C_{2}\parallel M_z]$ relates the opposite spin sector at the opposite axial momentum,
\begin{equation}
E_\sigma(k_z)=E_{-\sigma}(-k_z).
\label{eq:S_tubesym}
\end{equation}
Combining Eqs.~\eqref{eq:S_spinonly} and \eqref{eq:S_tubesym} yields
\begin{equation}
E_\sigma(k_z)=E_{-\sigma}(k_z),
\label{eq:S_spindeg}
\end{equation}
}
consistent with a compensated collinear weak-SOC reference state. This axial degeneracy coexists with the spin-resolved helical character of the corresponding wave functions. \textcolor{black}{Equation~\eqref{eq:S_spindeg} relates the spectra as sets and may permute band indices. Symmetry fixes the spin parity of the axial--azimuthal response, while transport determines its magnitude.}

\subsection{\texorpdfstring{\textcolor{black}{Direct effect: spin-selective helical sheet current}}{Direct effect: spin-selective helical sheet current}}\label{subsec:direct}
In the direct configuration, \textcolor{black}{spin-selective injection into one fixed spin channel} drives a steady-state current density that develops a helical pattern whose handedness is locked to the spin orientation, as sketched in Figs.~\ref{figure1}(c,d). The helical charge flow produces an axial magnetic flux whose sign depends on the injected spin.

Let $z$, $\theta$, $R$, and $t_{\rm w}$ denote the tube axis, azimuthal coordinate, mean radius, and wall thickness. The spin-resolved sheet current is
{\color{black}
\begin{equation}
\boldsymbol{\mathcal K}^{\sigma}(z,\theta)
\equiv
\int_{R-t_{\rm w}/2}^{R+t_{\rm w}/2}dr\,
\boldsymbol J^{\sigma}(r,z,\theta)
=
\mathcal K_z^{\sigma}\hat{\bm z}
+\mathcal K_\theta^{\sigma}\hat{\bm\theta},
\label{eq:sheet-current-definition}
\end{equation}
where $\boldsymbol J^{\sigma}$ is the volume current density in the wall and $\boldsymbol{\mathcal K}^{\sigma}$ has units of A\,m$^{-1}$.
}
For a uniform current, the chirality angle and helical pitch are
{\color{black}
\begin{equation}
\tan\alpha_\sigma
=\frac{\mathcal K_\theta^\sigma}{\mathcal K_z^\sigma},
\qquad
\Lambda_\sigma
=2\pi R\frac{\mathcal K_z^\sigma}{\mathcal K_\theta^\sigma}.
\label{eq:alphaLambda}
\end{equation}
}
The spin-space relation that exchanges the fixed spin channels preserves the mapped axial component and reverses the azimuthal component. For otherwise identical injections,
{\color{black}
\begin{equation}
\mathcal K_z^\uparrow=\mathcal K_z^\downarrow,
\qquad
\mathcal K_\theta^\uparrow=-\mathcal K_\theta^\downarrow,
\label{eq:helicitylock}
\end{equation}
}
so $\alpha_\uparrow=-\alpha_\downarrow$ on the principal branch and $\Lambda_\uparrow=-\Lambda_\downarrow$.

\textcolor{black}{For the rolled-sheet mapping with a common relaxation time, the parent-layer sheet-conductance tensor in Eq.~\eqref{eq:boltzmann-tensor} gives $\mathcal K_\theta^\sigma=\sigma_{\theta z}^\sigma E_z$, with $\sigma_{\theta z}^\uparrow=-\sigma_{\theta z}^\downarrow$ for the compensated weak-SOC spin sectors~\cite{Manc2021}.}
\textcolor{black}{For a long, uniformly conducting thin-walled tube and positions far from its ends, Amp\`ere's law gives}
{\color{black}
\begin{equation}
B_{z,\mathrm{in}}^{(\sigma)}
\simeq\mu_0\mathcal K_\theta^{\sigma},
\label{eq:axisfield}
\end{equation}
}
which reverses when the selected fixed spin channel is reversed.

Building on the microscopic helicity above, in the direct configuration, a spin imbalance writes a definite helical handedness into an otherwise achiral nanotube: the spin-dependent azimuthal response generates a surface current and an axial near field, both reversing upon reversal of the spin. This magnetoelectric texture breaks mirror symmetry in the local environment. Although originating in transport, it provides a symmetry-allowed channel by which spin polarization can be converted into an emergent chirality of the interfacial potential and of the spin-polarized density of states sensed by adsorbates.

Thermally activated carriers near the spin-polarized band edge therefore experience a spin-selective angular-momentum bias, forming circular charge and spin currents that couple to molecular orbitals via exchange and magnetic-dipole interactions. As a consequence, reaction intermediates adsorbed on the tube wall encounter a spin-dependent potential landscape, potentially favoring one enantiomeric pathway over its mirror counterpart.

Two-dimensional altermagnetic materials are still scarce, and the catalytic scenarios to which the above properties can be applied are correspondingly limited. Nevertheless, as shown in Fig.~\ref{figure1}(e), assembling coaxial bilayer heterostructure nanotubes allows the altermagnetic tube to imprint its spin-selected helicity onto a nominally achiral inner or outer tube via electromagnetic and exchange/tunneling couplings, with the induced handedness reversing when the spin is reversed. In this division of roles, the altermagnetic tube programs chirality while the partner tube provides the catalytic environment, yielding a more general spin-tunable chiral nanotube catalyst. Because the control parameter is spin rather than static lattice asymmetry, the handedness is fully programmable.

\subsection{\texorpdfstring{\textcolor{black}{Inverse effect: quasistatic flux-driven spin-channel conversion}}{Inverse effect: quasistatic flux-driven spin-channel conversion}}\label{subsec:inverse}
\textcolor{black}{The inverse response is considered in the quasistatic intraband regime, $\omega\tau\ll1$, at frequencies below the relevant interband scales.}
The inverse configuration is shown in Fig.~\ref{figure1}(f). We take $e>0$ for notational clarity. A time-varying axial flux $\Phi(t)$ induces a circumferential Faraday field,
\begin{equation}
\oint \mathbf{E}\!\cdot\! d\boldsymbol{\ell}=-\dot{\Phi}
\quad\Rightarrow\quad
E_\theta(r)=-\frac{\dot{\Phi}}{2\pi r},
\label{eq:faraday}
\end{equation}
which is approximately uniform through a thin wall, $t_{\rm w}\ll R$, so $E_\theta(r)\simeq E_\theta(R)$.

In a rolled altermagnet, the same spin-space-symmetry argument allows an axial--azimuthal cross response with opposite signs for the two fixed spin channels. \textcolor{black}{Combining the parent-layer cross conductivity with the standard two-current electrochemical-potential form~\cite{Manc2021,ValetFert1993} gives}
{\color{black}
\begin{equation}
\mathcal K_z^\sigma(z)
=\sigma_{z\theta}^\sigma E_\theta(R)
-\frac{\sigma_{zz}^\sigma}{e}\,\partial_z\mu_\sigma(z),
\label{eq:Kzsigma}
\end{equation}
}
where $\mu_\sigma$ is the spin-resolved electrochemical energy. \textcolor{black}{For the intraband Boltzmann tensor used here, $\sigma_{z\theta}^\sigma=\sigma_{\theta z}^\sigma$.} \textcolor{black}{For compensated common-$\tau$ weak-SOC spin sectors, the two channels obey~\cite{Manc2021}}
\begin{equation}
\sigma_{z\theta}^\uparrow=-\sigma_{z\theta}^\downarrow,\qquad
\sigma_{zz}^\uparrow=\sigma_{zz}^\downarrow\equiv\sigma_0 .
\label{eq:AMsym}
\end{equation}

Following the standard two-current convention~\cite{ValetFert1993}, define the charge and spin electrochemical energies as
{\color{black}
\begin{equation}
\mu_c=\frac{\mu_\uparrow+\mu_\downarrow}{2},\qquad
\mu_s=\frac{\mu_\uparrow-\mu_\downarrow}{2},
\end{equation}
}
and the axial charge and spin-channel sheet currents as
{\color{black}
\begin{equation}
\mathcal K_z^c=\mathcal K_z^\uparrow+\mathcal K_z^\downarrow,\qquad
\mathcal K_z^s=\frac{\hbar}{2e}
\left(\mathcal K_z^\uparrow-\mathcal K_z^\downarrow\right).
\end{equation}
}
Using Eqs.~\eqref{eq:Kzsigma} and \eqref{eq:AMsym} gives
{\color{black}
\begin{align}
\mathcal K_z^c
&=-\frac{2\sigma_0}{e}\,\partial_z\mu_c,
\label{eq:Kc}\\
\mathcal K_z^s
&=\frac{\hbar}{e}\sigma_{z\theta}^\uparrow E_\theta(R)
-\frac{\hbar\sigma_0}{e^2}\,\partial_z\mu_s.
\label{eq:Ks}
\end{align}
}
\textcolor{black}{A two-terminal zero-bias measurement imposes $I_c=2\pi R\mathcal K_z^c=0$. For a uniform tube, Eq.~\eqref{eq:Kc} then gives $\partial_z\mu_c=0$. Far from the contacts of a long homogeneous tube, we further take $\partial_z\mu_s\simeq0$.}
Combining Eqs.~\eqref{eq:faraday} and \eqref{eq:Ks} yields
{\color{black}
\begin{equation}
\mathcal K_z^c=0,\qquad
\mathcal K_z^s
=-\frac{\hbar}{2\pi eR}
\sigma_{z\theta}^\uparrow\dot{\Phi}.
\label{eq:purespin-bulk}
\end{equation}
}
\textcolor{black}{Thus, the Faraday field generates a charge-neutral axial spin-channel sheet current in the tube bulk.}

\subsection{\texorpdfstring{V$_2$Se$_2$O nanotube calculations}{V2Se2O nanotube calculations}}\label{subsec:v2seo}
\begin{figure}[htbp]
\includegraphics[scale=1]{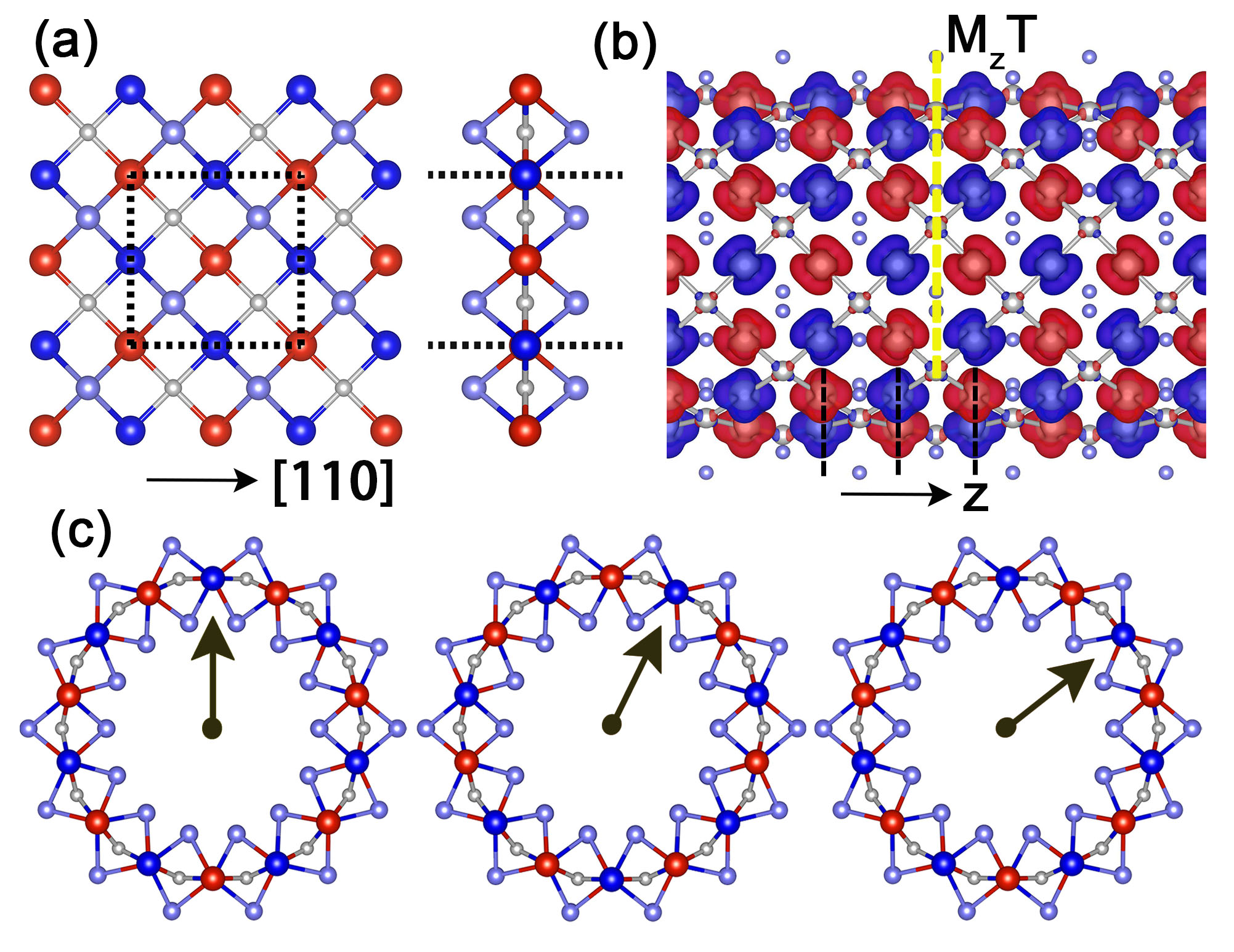}
\caption{
(a) Atomic structure of two-dimensional V$_2$Se$_2$O. Red and blue spheres denote V atoms with opposite spin orientations, gray spheres are O atoms, and light-purple spheres are Se atoms. The [110]-direction is shown by the black arrow.
(b) Spin-polarized charge density of a V$_2$Se$_2$O nanotube, showing the \textcolor{black}{spin-resolved helical modulation along the tube axis $z$}. \textcolor{black}{The yellow line marks a lattice mirror plane.}
(c) Cross-sectional views corresponding to three axial planes in (b), arrows highlight the helical correlation between the spin-polarized lobes around the tube circumference.
}
\label{figure2}
\end{figure}

Using V$_2$Se$_2$O as an illustrative example, the monolayer structure is shown in Fig.~\ref{figure2}(a). When rolled into a \textcolor{black}{relaxed [110], $N=7$ nanotube} containing 14 parent unit cells \textcolor{black}{with an axial period of 5.772~\AA}, the spin-polarized charge density [Fig.~\ref{figure2}(b)] reveals a distinct spin texture: the spin-up (red) and spin-down (blue) charge lobes form counter-rotating helical patterns along the tube axis $z$. The three successive cross-sectional slices in Fig.~\ref{figure2}(c) demonstrate the continuous rotation of the spin-polarized charge density around the tube circumference.

\begin{figure}[htbp]
\includegraphics[scale=1]{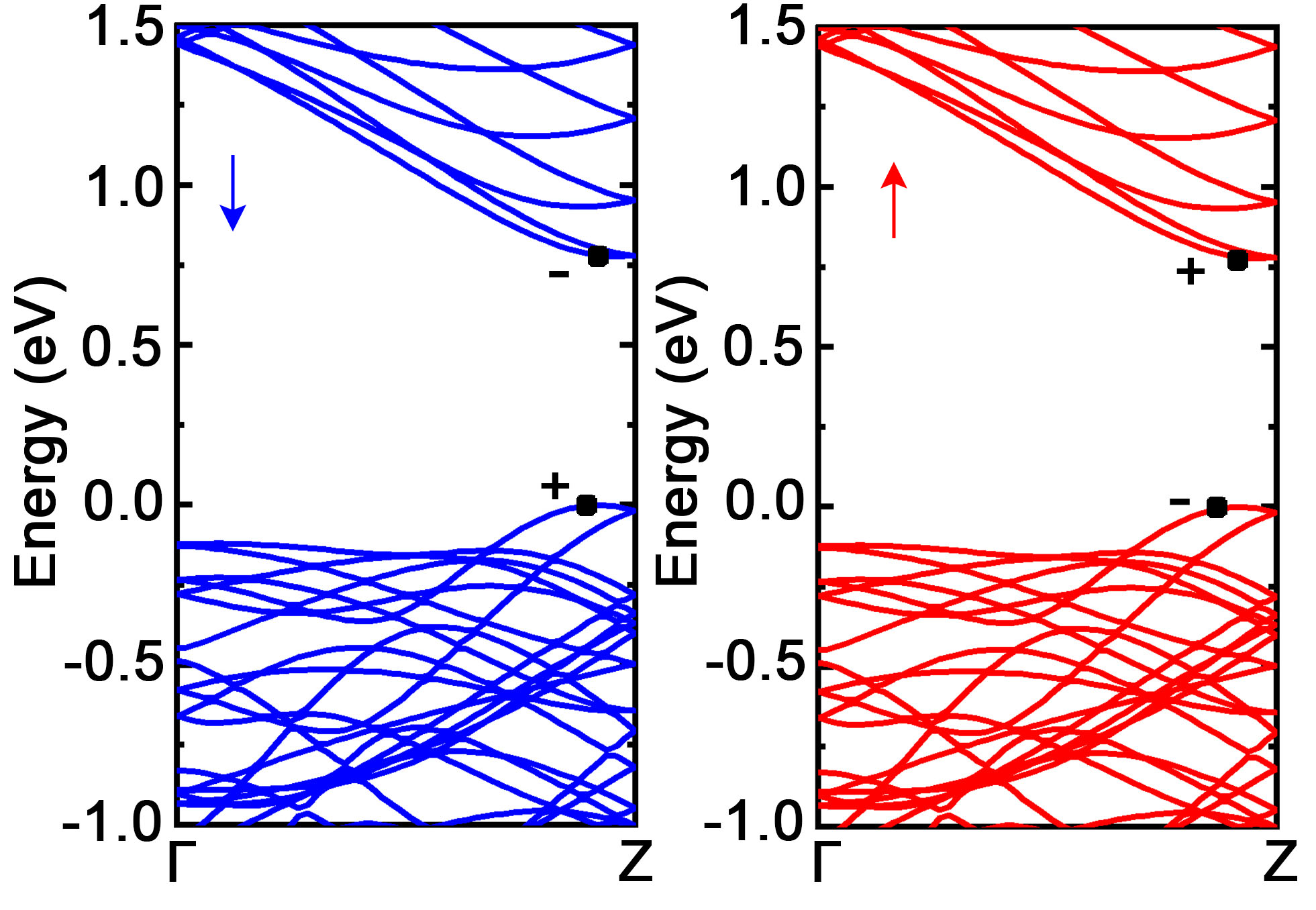}
\caption{Spin-resolved electronic band structures of the V$_2$Se$_2$O nanotube. Red and blue lines denote spin-up and spin-down channels, respectively. The CBM and VBM are indicated by black squares, where the symbols $+$ and $-$ label clockwise and anticlockwise helical wave-function characters along the $z$ axis.
}
\label{figure3}
\end{figure}

\begin{figure}[htbp]
\includegraphics[scale=1]{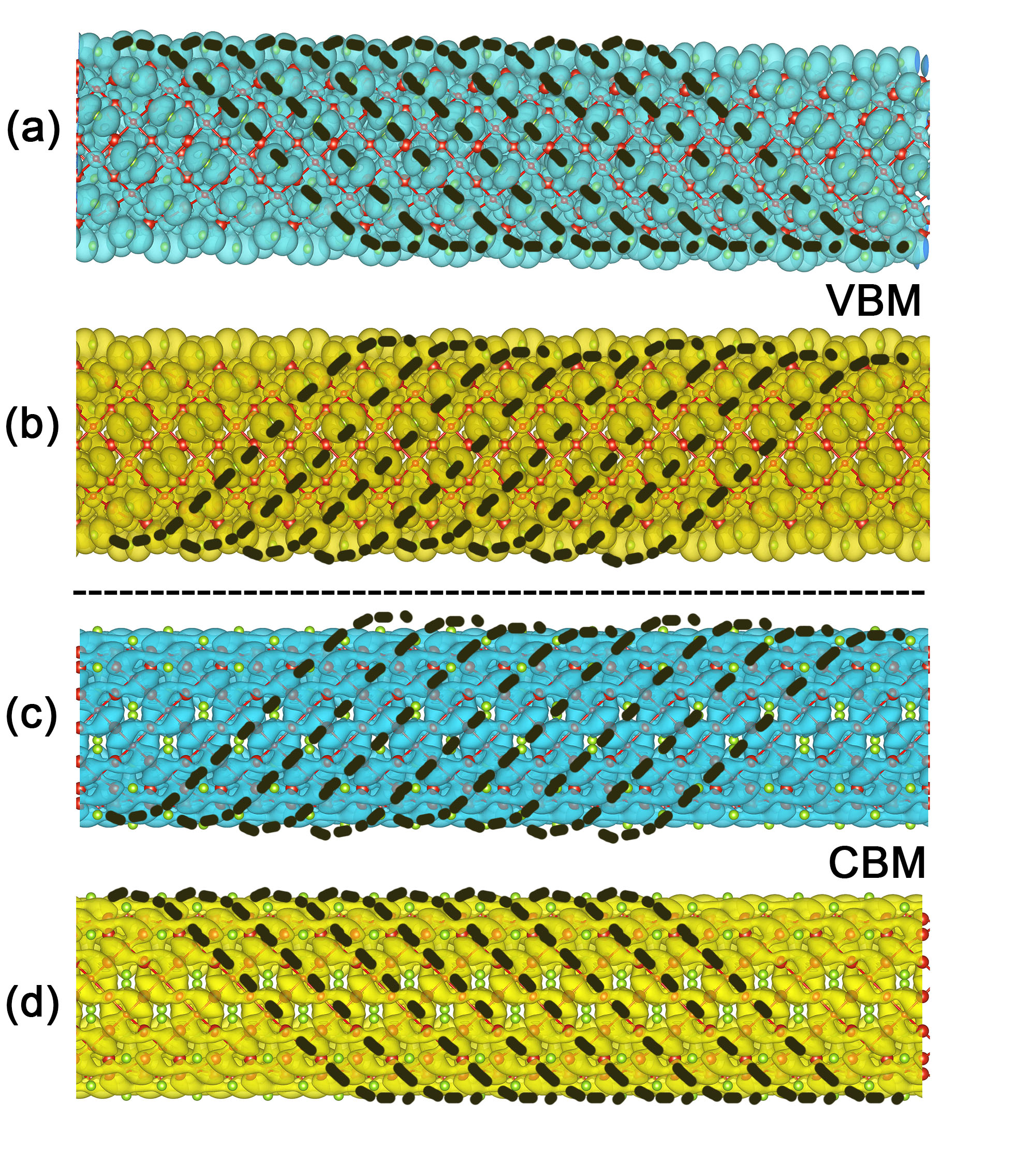}
\caption{
Real-space charge densities (squared modulus of the wave functions) of the VBM and CBM states for the V$_2$Se$_2$O nanotube, corresponding to the black points in Fig. \ref{figure3}. Panels (a,c) show the spin-down (blue) channel, and panels (b,d) show the spin-up (yellow) channel.
}
\label{figure4}
\end{figure}

The corresponding spin-resolved electronic band structures are shown in Fig.~\ref{figure3}. The spin-up and spin-down channels, plotted in red and blue, exhibit nearly degenerate dispersions along the $z$ direction and opposite helical wave-function characters. The conduction-band minimum (CBM) and valence-band maximum (VBM) occur at the marked $k_z$ points. The symbols $+$ and $-$ identify clockwise and anticlockwise helical wave-function characters visualized in Fig.~\ref{figure4}.

The real-space character of these band-edge states is illustrated in Fig.~\ref{figure4}, which plots the squared wave functions of the spin-down and spin-up channels for the VBM and CBM, respectively. Both states display pronounced helical charge-density modulations along the tube axis, consistent with the spin-dependent helicities identified in Fig.~\ref{figure3}. This spin--chirality locking is a purely nonrelativistic effect arising from the interplay between altermagnetic spin polarization in momentum space and the screw symmetry of the rolled geometry.

\begin{figure}[htbp]
\centering
\includegraphics[width=8.3cm]{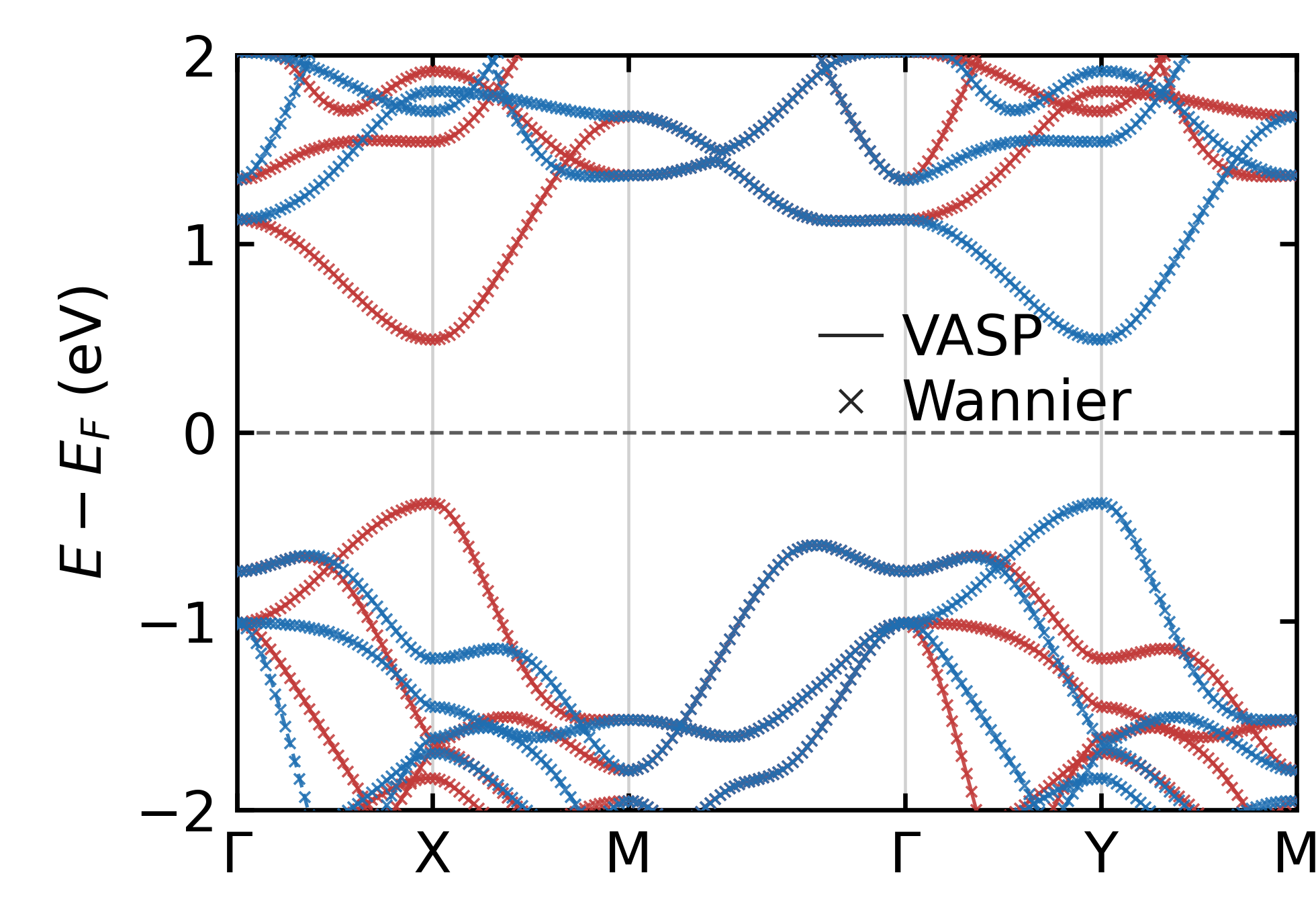}
\caption{\textcolor{black}{Parent-layer V$_2$Se$_2$O DFT--Wannier band comparison. Solid lines and crosses denote VASP and Wannier energies, respectively. Red and blue identify the two fixed spin channels.}}
\label{fig:v2-wannier-fit}
\end{figure}

\label{subsec:parent-wannier}
{\color{black}
Figure~\ref{fig:v2-wannier-fit} compares the parent-layer DFT bands with the 19-function-per-spin Wannier interpolation. Their agreement benchmarks the near-edge energy interpolation from which the velocities in Eq.~\eqref{eq:wannier-velocity} are evaluated.
}

\label{subsec:scope}
{\color{black}
Ma \textit{et al.} reported an equivalent spin Hall angle of approximately \(0.7\) for monolayer V$_2$Se$_2$O at \(\mu-E_{\rm VBM}=-0.20\)~eV when the electric field is applied along [110]~\cite{Manc2021}. In our spin-resolved conductivity convention, this observable is the ratio of the transverse spin conductivity to the longitudinal charge conductivity,
\begin{equation}
|S_{\pi/4}|
=\left|\frac{\sigma_{\theta z}^{\uparrow}-\sigma_{\theta z}^{\downarrow}}
{\sigma_{zz}^{\uparrow}+\sigma_{zz}^{\downarrow}}\right|\simeq0.7 .
\end{equation}

{\color{black}
At the same reference energy and transport orientation, our 19-Wannier-function-per-spin parent Hamiltonian, evaluated at \(T=150\)~K, gives $\eta_{\uparrow}=-0.651$ and $\eta_{\downarrow}=+0.651$.

The numerical residual, $|\eta_{\rm charge}|=1.1\times10^{-6}$, is consistent with the symmetry-expected cancellation between the two spin sectors. Because the same spin-independent relaxation time \(\tau\) is used for all conductivity components, it cancels from \(S_{\pi/4}\), \(\eta_{\rm spin}\), and \(\eta_{\rm charge}\), no numerical value of \(\tau\) is required for these dimensionless ratios.

The effective masses are obtained from three-point band-edge curvatures only to connect the full-Brillouin-zone calculation to the analytic parabolic-valley model of Ref.~\cite{Manc2021} and the dashed band-edge limits in Fig.~\ref{fig:v2-parent-transport-energy}. At the $X$ valley, the resulting hole masses are $m_{a_1}^{*}=1.508\,m_e$ and $m_{a_2}^{*}=0.294\,m_e$, where $m_e$ is the free-electron mass; the components are interchanged at the symmetry-related $Y$ valley. For the [110] geometry, the parabolic model reduces to $|S_{\pi/4}^{\rm par}|=|m_{a_1}^{*}-m_{a_2}^{*}|/(m_{a_1}^{*}+m_{a_2}^{*})=0.673$. This curvature-based value is only the band-edge limit, whereas $|\eta_{\rm spin}|=0.651$ is obtained directly from the full-Brillouin-zone Wannier--Boltzmann integral at $\mu-E_{\rm VBM}=-0.20$~eV. The difference reflects nonparabolic full-band contributions.
}

{\color{black}
Figure~\ref{fig:v2-parent-transport-energy} resolves the carrier-energy dependence of the parent-layer response. For $+E_z$, we choose $+\hat{\mathbf z}\parallel\mathbf a_1+\mathbf a_2$ and $+\hat{\boldsymbol\theta}\parallel\mathbf a_1-\mathbf a_2$. Under this convention, $\eta_\uparrow<0$ and $\eta_\downarrow>0$, so their transverse charge-current components point along $-\hat{\boldsymbol\theta}$ and $+\hat{\boldsymbol\theta}$, respectively.

In the compensated limit, $\eta_\uparrow=-\eta_\downarrow=\eta_{\rm spin}$. For the ideal-rolling estimate, define the total azimuthal sheet current $\mathcal K_\theta=\mathcal K_\theta^\uparrow+\mathcal K_\theta^\downarrow$, the signed axial current difference $\Delta I_z=I_z^\uparrow-I_z^\downarrow$, and $I_z^{\rm pol}=|\Delta I_z|$. Distributing the axial currents uniformly around a tube of radius $R$ gives $\mathcal K_z^\sigma=I_z^\sigma/(2\pi R)$. The parent-layer conversion and Amp\`ere's law for a long, uniformly conducting thin wall give
\begin{align}
\mathcal K_\theta(\mu)
&=\eta_{\rm spin}(\mu)\frac{\Delta I_z}{2\pi R},\nonumber\\
|B_{z,\mathrm{in}}(\mu)|
&=\mu_0|\mathcal K_\theta(\mu)|
=\frac{\mu_0|\eta_{\rm spin}(\mu)|I_z^{\rm pol}}{2\pi R}.
\label{eq:energy-dependent-field}
\end{align}
For $I_z^{\rm pol}=1~\mu\mathrm{A}$ and $R=0.64$~nm, the two reference points give $203~\mu\mathrm{T}$ and $115~\mu\mathrm{T}$. Panels~\ref{fig:v2-parent-transport-energy}(c) and (d) show the corresponding field scales for these parameters. Equal co-directed channel currents give $I_z^{\rm pol}=0$ and hence no interior axial field.
}
\begin{figure}[!t]
\centering
\includegraphics[width=8.3cm]{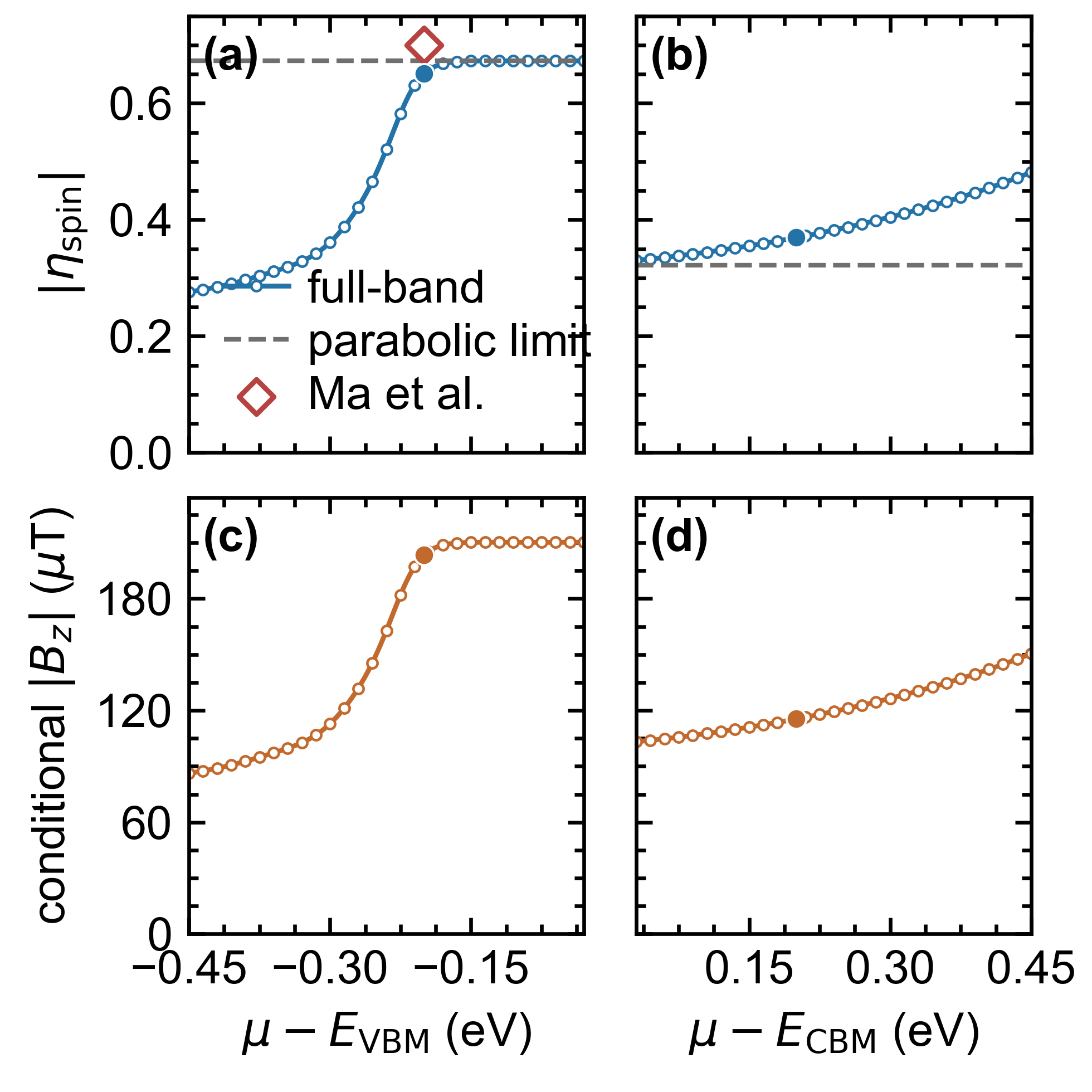}
\caption{\textcolor{black}{Energy-dependent spin conversion and ideal thin-wall field scale for monolayer V$_2$Se$_2$O. Panels (a) and (b) show $|\eta_{\rm spin}(\mu)|$ on the hole- and electron-doped sides from the 150~K Wannier--Boltzmann calculation. Open circles connected by solid lines show the full-band calculations, dashed lines the band-edge effective-mass limits, filled circles the $\pm0.20$~eV reference points, and the open diamond the value reported by Ma \textit{et al.}~\cite{Manc2021}. Panels (c) and (d) show the corresponding $|B_{z,\mathrm{in}}|$ for $I_z^{\rm pol}=1~\mu\mathrm{A}$ and $R=0.64$~nm.}}
\label{fig:v2-parent-transport-energy}
\end{figure}

{\color{black}
The explicit nanotube DFT calculation establishes the axial bands and spin-resolved band-edge $|\psi|^2$ modulation, while the parent-layer Wannier--Boltzmann calculation quantifies the rotated two-dimensional conductivity tensor.
The field values are ideal thin-wall estimates. Quantitative finite-tube response requires a relaxed-tube transport calculation.
Reports of single-walled MoS$_2$ nanotubes near the 1-nm scale demonstrate the feasibility of ultrathin TMD tubes~\cite{Nakanishi2026ScienceMoS2NT}. Possible tests include local magnetometry under spin-selective axial injection and spin-sensitive detection under an axial ac flux.
}
}

Relatedly, the recently reported spin-axis dynamic locking (SADL) \cite{zhiheng2025spinaxisdynamiclocking,lai2025dwaveflatfermisurface} in two-dimensional altermagnets shows that the direction of the driving electric field determines the spin polarization of the resulting current: an electric field along one crystal axis generates a purely spin-up current, while along the orthogonal axis it produces a purely spin-down current. A diagonal field can even achieve nearly 100\% charge-to-spin conversion. This effect originates from the same sublattice-anisotropy mechanisms discussed here. When such SADL-active layers are rolled into nanotubes, their principal axes naturally align with the tube's axial and circumferential directions, which is expected to further strengthen the coupling between spin and helical charge flow, leading to an enhanced spin-dependent chiral response.

\section{Summary}
In summary, rolling a two-dimensional altermagnet into a nanotube converts its \textcolor{black}{directional spin anisotropy in momentum space} into a spin--chirality coupling enforced by screw symmetry. This yields \textcolor{black}{two reciprocal effects}: (i) \textcolor{black}{spin-selective} injection produces a helical current whose handedness is locked to the spin, generating opposite-sign axial magnetic fields, and (ii) a time-varying axial flux drives equal-magnitude, opposite-sign axial charge currents in the two \textcolor{black}{fixed} spin channels, \textcolor{black}{yielding zero net axial charge current but a finite spin-channel current}. \textcolor{black}{Relaxed-tube DFT reveals spin-resolved helical $|\psi|^2$ modulations at the band edges, while parent-layer Wannier--Boltzmann transport quantifies the transverse response, with $|\eta_{\rm spin}|\simeq0.65$ near the hole reference point. Together, these results establish the rolled-geometry mechanism and quantify its parent-layer response.}

\begin{acknowledgments}
We would like to thank Shandong Institute of Advanced Technology and National Supercomputing Center (Shuguang) for providing computational resources. This work is supported by the National Natural Science Foundation of Shandong Province(Grant No. ZR2024QA040), Natural Science Foundation of Hebei Province (Grant No.A2025202032). Xin Chen thanks China scholarship council for financial support (No. 201606220031).
\end{acknowledgments}

\bibliography{Ref}% Produces the bibliography via BibTeX.

\end{document}